\documentclass{article}
\usepackage{cite}
\usepackage{graphicx}
\usepackage{amssymb}
\usepackage{authblk}
\usepackage{url}
\usepackage{amsmath}

\title{A Potts Model for Night Light \\and Human Population}
\author[1]{Gabriell~M\'at\'e \thanks{Corresponding author\\email: g.mate@tphys.uni-heidelberg.de}}

\affil[1]{Institute for Theoretical Physics, Heidelberg University, Philosophenweg 19, Heidelberg, Germany}

\date{\today}

\begin{document}

\maketitle

\begin{abstract}
The Potts model was one of the most popular physics models of the twentieth century in an interdisciplinary context. It has been applied to a large variety of problems.
 Many generalizations exists and a whole range of models were inspired by this statistical physics tool. Here we present how a generic Potts model can be used to study
complex data. As a demonstration, we engage our model in the analysis of night light patterns of human settlements observed on space photographs.
\end{abstract}

%
\section{Introduction}
%

The Potts model \cite{potts_generalized_1952} has enjoyed a great popularity in the second half of the last century. As a generalization of the Ising model \cite{ising_beitrag_1925}, it not only explained different aspects of ferromagnetic systems, but it was also successfully used for studying interdisciplinary phenomena. Examples include community structure detection, tumor growth and especially image segmentation \cite{reichardt_detecting_2004,rejniak_hybrid_2011,chen_image_2005}. In fact, many Image Processing and Machine Learning applications rely heavily on approaches similar to the Potts model. Inspired by Statistical Physics models, a whole model-family emerged. Nowadays, these are known as Undirected Graphical Models or Markov Random Fields (MRFs) \cite{blake_markov_2011}. In this sense, most of the MRF models can be viewed as a generalization of the Potts model.

In the generic framework of MRFs, many algorithms have been developed to perform a variety of tasks. Generally, a model with some parameters would be set up to describe a some empirical data. Then, usually the first question considers the parameter values for which the proposed model would capture the often complex nature of the data. To solve this tasks, specialized algorithms are used to calculate or approximate these parameters (see for instance \cite{redner_mixture_1984,manjunath_unsupervised_1991,boykov_markov_1998,descombes_estimation_1999,lafferty_conditional_2001,huang_generalized_2002}).

As a consequence of the availability of robust estimation approaches, MRFs are mostly used to represent complex dependency structures in random variables which describe some data, and to manipulate the data (for instance, detect clusters) based on the obtained representations. However, in most practical applications, the mathematical formalisms are rather abstract and the physical meanings of the estimated parameters are lost or not of interest. 

Our main goal in this study is to illustrate how generalized Potts models, that is MRFs, can be used to model data with complex interdependencies and how to gain access to hidden information by looking at the parameters of the model. Concept-wise, this approach is similar to a scenario in which we are given a sample of an ensemble generated by a Potts model at an unknown temperature $T$. In this case, we can estimate the temperature $T$ using the provided samples and thus characterize the system to the best of our knowledge.

Starting out from the original Potts model, first we will rewrite the model Hamiltonian, and transform it into a more general form. Then, we present the applicability of the idea through a concrete example. We will implement the rewritten model for the complex patterns of human settlements observed through night-light photographs of the Earth. We will estimate the parameters of the model, interpret them in the context of the data and, finally, based on the estimated values, we will draw conclusions regarding the spatial structure of the night light and settlement patterns.

\section{The Used Model}

\subsection{The Potts model in general}

In its original form, the Potts model is defined in the following way: A spin, encoded by a scalar variable with $q$ possible values, for instance, the integers $\{1, \ldots, q\}$, is placed on each site of  a (usually) regular lattice. These scalar variables represent the direction of the spins. The spins interact with their nearest neighbors and an external magnetic field $h'$ (if present). The interaction energy is defined by the Hamiltonian
\begin{equation}
	H = - J' \sum_{\langle i,j \rangle} \delta_{\sigma_i \sigma_j} - h' \sum_i \delta_{\sigma_i s} \mbox{,}
	\label{eq_potts_hamiltonian}
\end{equation}
where the $\langle i,j \rangle$ notation means neighboring $i$ and $j$ lattice indexes. $\sigma_i$ is the state of spin $i$, $J'$ is the coupling constant, defining the strength of the interaction between two neighboring spins found in similar states, $\delta$ is the Kronecker delta, and $s$ is a particular spin orientation. In this sense, the magnetic field is parallel with the spin orientation $s$.

Given a set of spin-configurations, known to be a sample of an ensemble generated by a Potts model at a given temperature $T$ with fixed parameters $J'$ and $h'$, it is rather straightforward to estimate the parameter combinations $J'/T$ and $h'/T$, and thus characterize the ensemble within the error limits set by the quality of the sample. This kind of estimation is called \textit{unsupervised learning}, as it requires no input (such as hand-labeling certain features in the samples) from humans.  

%
\subsection{The modified model}
%

While many modifications and extensions of the Potts model have been studied and engaged in different applications, here we will present a generalization of the model based on a relatively simple observation. Note that the Kronecker delta in the first term in Equation (\ref{eq_potts_hamiltonian}) is nothing else but a $q \times q$ unity matrix indexed by the states $\sigma_i$ and $\sigma_j$. According to this formalism, spins interact only if they are in the same state, that is, if they are parallel. Replacing this unity matrix by an arbitrary $q \times q$ symmetric matrix introduces interactions between non-parallel states. Furthermore, we can also rewrite the last term in the Hamiltonian, introducing an arbitrary interaction of the different states with the external magnetic field in the following way:
\begin{equation}
	H = - \sum_{\langle i,j \rangle} J_{\sigma_i \sigma_j} - \sum_i h_{\sigma_i} \mbox{,}
	\label{eq_potts_mod1_hamiltonian}
\end{equation}
where $J$ is a $q \times q$ symmetric matrix, and $h$ is a $q$ dimensional vector. $J$ here defines the strength of the interactions between the different orientations. For instance $J_{12}$ ($=J_{21}$) tells us how strongly spin orientations 1 and 2 interact if they are neighbors. Similarly, $J_{33}$ drives the interactions between two neighboring spins when they both are in state 3.

In a further step, we note that it is possible to imagine a locally changing external field with $r$ possible orientations. Assuming that this field may have different orientations around the different spins, and that it may interact with each spin orientation differently, the Hamiltonian can be extended as 
\begin{equation}
	H = - \sum_{\langle i,j \rangle} J_{\sigma_i \sigma_j} - \sum_i h_{\sigma_i} - \sum_{i} K'_{\sigma_i l_i}\mbox{,}
	\label{eq_potts_mod2_hamiltonian}
\end{equation}
where $l_i$ is the orientation of the locally changing external field around spin $i$, and $K'$ is a $q \times r$ matrix which defines the interactions of the spins with the locally changing field. Note that here, the orientations of the local field are also encoded by integers, in this case $l_i \in \{1, \ldots, r\}$. As an example, $K'_{21}$ tells us how strongly a spin in orientation 2 interacts with a locally changing field with an orientation {1}. This is not equivalent with $K'_{12}$ which tells us how a spin in state 1 interacts with a field oriented in the direction 2. Similarly, $K'_{44}$ dictates the interaction of spin state 4 with field orientation 4.

As a last generalization step, we observe that the  $h$ from the second sum in Equation (\ref{eq_potts_mod2_hamiltonian}) can be absorbed into $K'$ by adding $h_s$ to each entry in row $s$ of $K'$. As a result, we get 
\begin{equation}
	H = - \sum_{\langle i,j \rangle} J_{\sigma_i \sigma_j} - \sum_{i} K_{\sigma_i l_i}\mbox{.}
	\label{eq_hamiltonian}
\end{equation}
While the model in Equation (\ref{eq_hamiltonian}) is formally very similar to the original Potts model, it is also very general and flexible because of the matrix parameters. We can instantly get back the original Potts model from Equation (\ref{eq_potts_hamiltonian}) by replacing $J$ with $J' \mathbb{I}$ and $K$ by $h' L_s$, where $\mathbb{I}$ is the identity matrix and $L_s$ is a matrix with all zeros except the entry in the $s$-th row and $s$-th column, which is 1. 

Once the Hamiltonian is defined, the distribution over the configurations is given by the Boltzmann distribution:
\begin{equation}
	p(H|J,K) = \frac{1}{Z} e^{-H} \mbox{,}
	\label{eq_boltzmann_0}
\end{equation}
where $Z$ is the partition function which normalizes the distribution. Note that here the inverse temperature was absorbed into the parameters $J$ and $K$, thus when estimating the parameters, we in fact estimate the $J/\beta$ and $K/\beta$ ratios. Note also that in fact $H$ depends on the particular configuration $W$ of the spins and that of the locally changing external field, which will be denoted by $D$. This means that in some sense $H$ is a function of these configurations: ${H = H(W, D)}$.

\subsection{Learning \label{learning_section}}

As we mentioned, the main question in most applications of MRFs is the value of the model parameters, given some sample configuration. However, because the partition function $Z$ in Equation (\ref{eq_boltzmann_0}) contains exponentially many terms, the calculation of the likelihood function is computationally intractable for systems with many spins, therefore standard maximal likelihood estimations are not viable. However, there are plenty of alternatives.

Even though the likelihood itself cannot be calculated, learning approaches usually take the derivative of the log-likelihood as a common starting point. This derivative for the model defined in Equation (\ref{eq_hamiltonian}) can be given as
\begin{align}
        \frac{\partial \log P(\theta | \{S_0\})}{\partial \theta} &= \frac{1}{\partial \theta} \left[ \sum_{<i,j>} J_{\sigma_i \sigma_j} + \sum_i K_{\sigma_i l_i} - \log Z\right] \nonumber \\
         &= \phi_\theta (S_0) - \frac{\partial \log Z}{\partial \theta}\mbox{,}
         \label{derivative_log_likelihood}
\end{align}
where $\theta \in \left\{J_{ab} | a,b \in \{1,2,\ldots, q\}\right\} \cup \left\{K_{ac} | a \in \{1,2,\ldots,q\}, c \in \{1,2,\ldots, r\}\right\}$ is one of the parameters of the model, $S_0$ is some observed data for which we want to estimate the parameters $\theta$ (these observations should contain configurations both for $W$ and $D$), and $\phi_\theta(S)$ is what the Machine Learning literature calls the \textit{potential} corresponding to the parameter $\theta$ in a system with configuration $S$. Note that these potentials are not potentials in a physical sense. The potential for a given $J_{ab}$ parameter is the negative derivative of the Hamiltonian with respect to this parameter, and it can be given as
\begin{equation}
        \phi_{J_{ab}} = \sum_{<i,j>} \delta_{a \sigma_i}\delta_{b \sigma_j} \mbox{.}
\end{equation}
Similarly
\begin{equation}
        \phi_{K_{ac}} = \sum_{i} \delta_{a \sigma_i} \delta_{c l_i}\mbox{.}
\end{equation}

The last term in the derivative (\ref{derivative_log_likelihood}) is in fact the theoretical expected value of the potentials:
\begin{equation}
        \frac{\log Z}{\partial \theta} = \frac{\sum_S \phi_\theta(S) e^{-\sum_{<i,j>} J_{\sigma_i \sigma_j} - \sum_i K_{\sigma_i l_i}}}{Z} = \sum_S \phi_\theta(S) p(S) = \langle \phi_\theta \rangle_m \mbox{,}
\end{equation}
where the notation $\langle \cdot \rangle_m$ is the theoretical (model) average . As we already discussed this, this term is intractable, and its calculation or approximation is the main problem in parameter estimations.

Different methods handle this problem with different approaches. Some methods calculate a pseudo-likelihood  \cite{ji_consistent_1996}, others, like Markov chain Monte-Carlo (MCMC) maximum likelihood estimation, use importance sampling to estimate the partition function \cite{neath_monte_2006}.

\subsubsection{Stochastic Approximation Procedure}

In the present study we applied a method called \textit{Stochastic Approximation Procedure} (SAP) as presented in \cite{salakhutdinov2009learning}. This approach uses MCMC sampling to estimate the model average $\langle \phi_\theta \rangle_m$. SAP has been found to work very well when implemented for Markov Random Fields.

Let us define a Markov process characterized by a transition probability $\pi(S_t~\rightarrow~S_{t+1})$ which satisfies the detailed balance condition
\begin{equation}
        p(S_t)\pi(S_t~\rightarrow~S_{t+1}) = p(S_{t+1})\pi(S_{t+1}~\rightarrow~S_{t}) \mbox{.}
\end{equation}
We initialize $M$ Markov chains with a random initial configuration $\{S_{t=1}^1,\allowbreak S_{t=1}^2,\allowbreak \ldots,\allowbreak S_{t=1}^M\}$. The parameters $\theta$ are also initialized with random values. We simulate the Markov chains and after each Monte-Carlo step we calculate the $\langle \phi_\theta(S^{t}) \rangle_{MC}$ average of the potentials over the Monte-Carlo samples. Based on these averages we update the parameters according to the update rule
\begin{equation}
        \theta = \theta + \eta \left[ \phi_\theta(S_0) - \langle \phi_\theta(S^{t}) \rangle_{MC} \right] \nonumber \mbox{,}
\end{equation}
where $\eta$ is the learning rate and it is decreased after each update. In case we have a set $\{S_0^1, S_0^1, \ldots, S_0^N\}$ of observed data sample, we can replace the $\phi_\theta(S_0)$ with the $\langle \phi_\theta(S_0) \rangle$ average calculated over the observed samples. In this case, the update rule is modified and it writes as
\begin{equation}
        \theta = \theta + \eta \left[ \langle \phi_\theta(S_0) \rangle - \langle \phi_\theta(S^{t}) \rangle_{MC} \right] \mbox{.}
        \label{sap_update}
\end{equation}


To calculate the $\pi(S_t \rightarrow S_{t+1})$ transition probabilities, we can use the simple Metropolis algorithm \cite{metropolis_equation_1953}.

\section{Describing the night-light distribution}

In the following we will present how data with relatively complex interdependencies can be modeled with the MRF presented in Equation (\ref{eq_hamiltonian}). For this, we will use a relatively low resolution gridded population density data \cite{gridded_pop_data_2014} and night-light data \cite{earth_earths_2009} for the year 2000. The latter one is practically containing space photographs of the earth taken during night, wherein the artificial light of human settlements is captured. Many studies relate these night light patterns to the economic development of the regions \cite{nordhaus_sharper_2014}. They indicate that the intensity and density of the night light can be viewed as an objective economic development index, especially for countries with a low-quality statistical systems. 

\begin{figure}
    \begin{center}
          \includegraphics[width=0.9\textwidth]{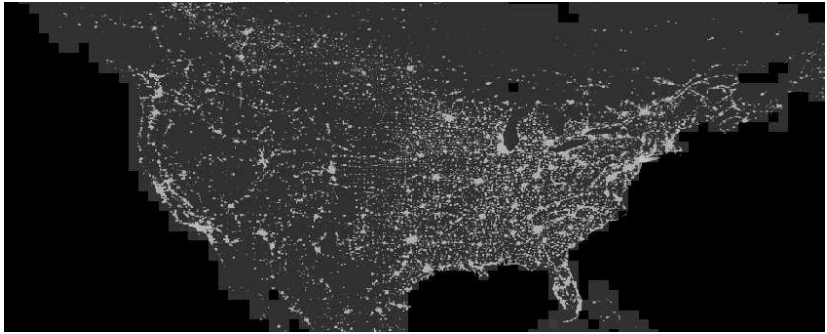}
          \caption{\label{fig:northam} Night light and population density coverage of North America. The a resolution of the night light data (bright patterns) is 1000x400, a cell corresponds to roughly 10 $kms$. The gray area indicates the geographical regions for which we have population density data. The side of a population density cell is approximately $120$ $kms$.}
    \end{center}
\end{figure}

We carried out the calculations using the data as is, that is, no corrections were applied. Without loss of generality, we chose to work only on North America as presented in Figure \ref{fig:northam}. First we matched the grid of the population density to that of the night light data. This meant refining the grid of the population density from a resolution of around $85 \times 35$ to a resolution of $1000 \times 400$. In this process no interpolation was used, we simply cut up the big population density cells to many small ones.

We implemented the model from Equation (\ref{eq_hamiltonian}) for the night light intensities $w$. In this case, the orientation (or the state) of a spins will correspond to a given intensity level of the night lights. On the other hand, we considered the population density $d$ as the external field, the environment which has an influence on the patterns in $w$. That is, we assumed the population density as constant, and thought about the night light patterns as feature resulting from the underlying spatial distribution of the population density. Since a spin in our model can point only to a limited number of directions, and similarly, the locally changing external field can have only a limited number of orientations, we discretized both the population density data and the night light intensity data. To achieve this, we need to project $w$ and $d$ to a discrete set. We will call the elements of this set states instead of orientations as this naming fits better in this context. In case of $w$, these states correspond to areas with weak, medium and strong night light (encoded with the integers 1,2 and 3, that is, $\sigma_i \in \{1,2,3\}$ for any $i$). Let us denote the sample of discretized light intensities with $W$. In the same fashion, for $d$, the discretized population density values indicate small, medium and large population densities (also encoded with the same integers, i.e. $l_i \in \{1,2,3\}$). The sample of discretized densities will be denoted by $D$.

\begin{figure}
    \begin{center}
          \includegraphics[width=0.5\textwidth]{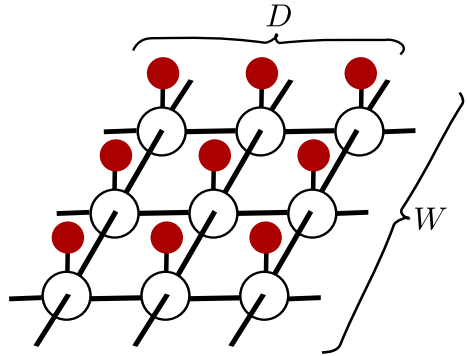}
          \caption{\label{fig:graphmod_for_light_and_pop} Graph representation of the MRF we implement for the example night light data. $W$ is the set of variables representing the discretized night light intensities while $D$ is the set of variables representing the discretized population densities.}
    \end{center}
\end{figure}

Because of the spatial structure of the data, we arranged the variables representing the night light according to a two dimensional square lattice, and, since there is known  “spillover effect” (meaning that new infrastructural/economic investments tend to be made next to already developed regions), we connected first order neighbors. Then we added another layer of variables which represent the population densities $d$. In this case, the graph representation of the model corresponds to the one in Figure \ref{fig:graphmod_for_light_and_pop}. Note however, that in other scenarios, variables might be arranged in structures other than a regular lattice. In fact, this approach allows the usage of arbitrary networks. Note also that variables representing the population densities are not connected to each other, that is they don't depend on each other in this study. We chose this implementation only because we focus on the night light patterns and consider the population densities as constants. As such, we did not care about how they depend on the densities in the neighboring regions. 

Again, the energy of the system is defined by Equation (\ref{eq_hamiltonian}). The values of $J$ will influence how compatible two different possible values of $\sigma$ are (for instance, $J_{11}$ will score how likely it is to observe a dim region next to another dim region, while $J_{23}$ drives the probability of finding a bright area next to a medium lit area). On the other hand, K defines the compatibility of the night light states with the states of the variables representing the population densities (the latter variable being  represented with red spheres in Figure \ref{fig:graphmod_for_light_and_pop}). For example, while $K_{12}$ indicates how likely it is to have a dark area coupled with a medium population density, $K_{31}$ will influence the probability of finding a bright region with a low population density. Let us emphasize, that the configuration of the locally changing external field is constant $D$, and in fact mathematically, it can be handled as a parameter of the distribution. Then, in the formalism of the Canonical Ensemble \cite{pathria_statistical_2011}, the probability distribution over the possible configurations in $W$ can be given as
\begin{equation}
	P(W|D,J,K) = \frac{1}{Z} e^{-H} \mbox{,}
	\label{eq_canonical_distribution}
\end{equation}
where again the inverse temperature term was absorbed in the parameters of the Hamiltonian $J$ and $K$. While $W$, $D$, $J$ and $K$ do not show up directly on the right hand side of Equation (\ref{eq_canonical_distribution}), the distribution is a function of these parameters/variables as the Hamiltonian $H$ depends on them. Another important thing to mention is that if the local Markov property is satisfied, meaning that a given spin probabilistically depends only on the variables that it is directly connected with in the graph representation, the Hammersley-Clifford theorem \cite{besag_spatial-temporal_1977} states that the probability distribution of a given night light configuration $W$ can always be given in the form Equation (\ref{eq_canonical_distribution}). This is a fundamental point which establishes a connection between different MRFs and ensures that algorithms developed for a given MRF can be adapted to other models of the family.

Using the sample data, presented in Figure \ref{fig:northam}, the model parameters $J$ and $K$ are then estimated through the approach presented in Section \ref{learning_section}\hspace{2pt}. The results of this estimation are presented in Table \ref{JK_results_table}. Note that the absolute values of the parameters do not matter, it is their relative values which is important. Therefore, in order to easily perceive the relative differences, the values in the matrices are shifted by subtracting the minimums of each $J$ and $K$ matrix from all of the entries of the corresponding matrix, thus the new minimums are 0. At this point, let us mention that, since the values of $J$ and $K$ are parameters playing similar roles in the same energy function, we can compare their values. This is one of the beneficial ``side effects'' of the approach presented here. Such a comparison would not be possible if we would simply calculate neighboring probabilities and population density and light intensity pairing probabilities, as the these  belong to different probability spaces.

\begin{table}[ht!]
        \centering
        \begin{tabular}{|c|c|c||c|c|c|}
          \hline
          \multicolumn{3}{|c||}{$J$} & \multicolumn{3}{|c|}{$K$} \\
          \hline
	  0.5316 & 0.3370 & 0 & 0.5851 & 0.5519 & 0.3780 \\
          \hline
	  0.3370 & 0.4570 & 0.2234 & 1.9544 & 2.0211 & 2.0554 \\
          \hline
	  0 & 0.2234 & 1.2354 & 0 & 0.6043 & 0.8941 \\
          \hline
        \end{tabular}
        \caption{\label{JK_results_table}Results for the estimation of the $J$ and $K$ parameters.}  
\end{table}

Analyzing the values in Table \ref{JK_results_table}, we can deduce the following: If we look at the spillover effect of the night light, which is given by J, the larger entries along the diagonal of J indicate that it is more probable to observe same intensity levels than different ones in neighboring cells. This effect is the strongest for high intensity regions. On the other hand, comparing the J values with the values in K, we see that the “prerequisite” for observing low light regions is driven both by having low light surrounding and low or at most medium density population (J11  K11 and K12 are comparable). At the same time, it is the least likely that we will find dark areas associated with high population densities (K13 is the smallest in row 1 of K). Although it is less salient when looking at the learned parameters, the situation is similar for medium lit regions in the sense that these regions will most probably have a medium lit surrounding and a medium or high population density, the latter is scored slightly better. These regions could correspond, for instance, to living neighborhoods, regardless whether they are located in a metropolis area or more in the country side. Finally, for bright regions it is much more likely to be observed close to other brightly lit regions, and this effect is more important than the population density in the area (J33 has the biggest weight compared to other entries in the third row of J and K). These territories most probably correspond to bright metropolitan centers.

\section{Discussion and Conclusions}
 
We presented how a generalized Potts model can be used to model and study data with complex spatial interdependencies. Starting out from the original Potts model, we rewrote the Hamiltonian allowing arbitrary interactions between any two spin-states. In addition, we considered a locally changing external field which may have a different orientation from spin to spin. We present the usefulness of the approach through a simple example: We investigated the interdependencies of the night light patterns and population densities of a given geographic territory. 

The introduced modifications of the Potts model enabled us to think about the night light as a spin configuration. In this case, the population densities were taken into account by considering them as a locally changing external field. Similarly to the interaction between the different spin-states, the effect of the locally changing filed on the different spin state was also considered arbitrary. In this setting, the aim is to estimate all these interactions. 

First, we presented how parameters can be estimated in such a framework. Then, after some necessary data preprocessing steps, which were kept minimal, we estimated the parameters. For this, we used an in-house developed software in our work. Once the parameters were learned based on the data sample, we were ready to draw our conclusions. Here we exploited the advantage of the presented framework, which consists in the fact that the parameters $J$ and $K$ are of the same nature. 

Based on such calculations, we could observe how territories with certain light intensity group while for the formation of other intensity levels the population density is a more important factor. Our simple conclusions make sense even if the data we used is rather imprecise as the resolution of the population density map we used is extremely coarse. It is obvious that with good quality data the outcomes would be more reliable and relevant.


\section{Acknowledgments}

The author would like to thank Dieter W. Heermann, Zolt\'an N\'eda and Miriam Fritsche for the useful discussions.

\bibliography{gabriell_mate_population_night_light}{}
\bibliographystyle{unsrt} 

\end{document}